\documentclass[10pt]{article}
\usepackage{base}
\usepackage{pxfonts}
\usepackage{tcolorbox}

\newcommand{\lp}[1]{\mathcal{L}[{#1}]}
\newcommand{\pair}[1]{\langle {#1}\rangle}

\input xy
\xyoption{all}

\title{Computational interpretations of classical reasoning: From the epsilon calculus to stateful programs}
\author{Thomas Powell}
\date{Preprint, \today}

\begin{document}

\maketitle

\begin{abstract}
The problem of giving a computational meaning to classical reasoning lies at the heart of logic. This article surveys three famous solutions to this problem - the epsilon calculus, modified realizability and the dialectica interpretation - and re-examines them from a modern perspective, with a particular emphasis on connections with algorithms and programming.
\end{abstract}

\section{Introduction}
\label{sec-intro}

This paper grew out of two talks I gave on the subject of proof interpretations: The first, at a conference on Hilbert's epsilon calculus at the University of Montpellier in 2015, and the second, at the Humboldt Kolleg workshop on the \emph{mathesis universalis} held in Como in 2017. Both of these events were notable in they brought together researchers from mathematics, computer science and philosophy, and as a consequence, speakers were faced with the challenge of making their ideas appealing for and comprehensible to each of these groups.

The latter workshop was particularly fascinating in its focus on the \emph{mathesis universalis}: The search for a universal mode of thought, or language, capable of capturing and connecting ideas from all of the sciences. It struck me then that part of my research has been driven, in a certain sense, by the desire to uncover universal characteristics behind the central objects of my own field of study - proof interpretations - and in particular to conceive of a language in which the algorithmic ideas which underlie them can be elegantly expressed. As such, my initial motivation when writing this article was to collect some of my own thoughts in this direction and put them down on paper.

In doing so, however, I realised that an article of this kind could also serve another purpose: To give an accessible and high level overview of a number of proof interpretations which play a central role in the history of logic but are often poorly understood outside of specialist areas of application. Proof interpretations are, by their very nature, highly syntactic objects, and gaining an understanding of how they actually \emph{work} when applied to a concrete proof tends to require a certain amount of hands-on experience. But perhaps a simple and informal case study tackled in parallel by several proof interpretations may form an accessible and insightful introduction to these techniques which would complement the many excellent textbooks on the subject?

I settled on an article which combines both of these goals. In the first main section I give an overview of three very well-known computational interpretations of classical logic: Hilbert's epsilon calculus, Kreisel's modified realizability together with the A translation, and finally G\"{o}del's Dialectica interpretation. My tactic is suppress many of the complicated definitions and to focus instead on a very simple example - the drinkers paradox - and sketch how each interpretation would deal with it in turn. In doing so I am able to highlight some of the key features which characterise each of the interpretations but are often invisible until one has acquired a working knowledge of how they are applied in practise. 

In the second part, I turn my attention to the relationship between the interpretations and core algorithmic ideas which underlie them. I begin with a general discussion on proof interpretations and the roles they play in modern proof theory, then focus on two ideas which feature in current research on program extraction: learning procedures and stateful programs. Both of these have been used by myself and others to characterise algorithms connected to classical reasoning, and together offer an illustration of how traditional proof theoretic techniques can be reinterpreted in a modern setting. Here I focus primarily on the Dialectica interpretation, and continue to use the drinkers paradox as a running example, in order to facilitate direction comparison with the earlier section. 

\subsection*{Prerequisites}

I have sought to capture the spirit of the \emph{mathesis universalis} in another way, by making this article accessible to as general an audience as possible. Nevertheless, I assume the reader is acquainted first-order logic and formal reasoning, and I expect a passing familiarity with the typed lambda calculus will make what follows a lot easier to read. Just in case, I mention a few crucial things here. The \emph{finite types} are defined inductively by the following clauses:
\begin{itemize}

\item $\NN$ is a type;

\item if $\rho$ and $\tau$ are types, $\rho\to\tau$ is the type of functions from $\rho$ to $\tau$.

\end{itemize}
Often is it convenient to talk about product types $\rho\times \tau$ in addition. Functions which take functions as an argument are called \emph{functionals}. 

System T is the well-known calculus of primitive recursive functionals in all finite types, whose terms consist of variables $x^\rho,y^\rho,z^\rho,\ldots$ for each type, symbols for zero $0:\NN$ and successor $s:\NN\to\NN$, allow the construction of new terms via lambda abstraction $\lambda x.t$ and application $t(s)$, and finally, contain recursors $R_\rho$ of each type, which allow the definition of primitive recursive functionals. Further details on System T can be found in e.g. \cite{AvFef(1998.0)}.

\subsection*{A note on terminology}

Proof interpretations are often inconsistently named and confused with one another. In this paper, I will use the term \emph{functional interpretation} to denote any proof interpretation which maps proofs to functionals of higher-type: Thus both realizability and the Dialectica interpretation are functional interpretations. The latter is also referred to as G\"{o}del's functional interpretation, and consequently often as just `the functional interpretation'. I am as guilty as anyone for propagating such confusion elsewhere, but here, since I discuss a number of proof interpretations, I will rigidly stick to the name \emph{Dialectica}.

\section{The drinkers paradox}
\label{sec-DP}

The running example throughout this paper will be a simple theorem of classical predicate logic, commonly known as the \emph{drinkers paradox}. This nickname apparently originates with Smullyan \cite{Smullyan(1978.0)}, and refers to the following popular formulation of the theorem in natural language:
\begin{quote}
In any pub, there is someone that, if they are drinking, then everyone in the pub is drinking.
\end{quote}
Of course, in pure logical terms, the drinkers paradox is nothing more than the simple first-order formula 
\begin{equation*}\exists n(P(n)\to \forall m P(m))\end{equation*}
where here, since we will primarily work in the setting of classical arithmetic, $P$ is assumed to be some decidable predicate over the natural numbers $\NN$. In fact, to make things slightly simpler for the functional interpretations (and more interesting for the epsilon calculus) we will study the following prenexation of the drinkers paradox: 
\begin{equation*}
\DP \quad : \quad \exists n\forall m (P(n)\to P(m)). 
\end{equation*}
which will henceforth be labelled $\DP$. 

The drinkers paradox is appealing for the proof theorist because it has a one-line proof in classical logic which doesn't give us any way to actually compute the `canonical drinker' $n$. It goes as follows: Either everyone in the pub is drinking, in which case we can set $n:=0$, or there is at least one person $m$ who is not drinker, in which case we set $n:=m$.

As such, any computational interpretation of classical logic has to have some way of dealing with the drinkers paradox, and as we will see, despite its apparent simplicity, $\DP$ illuminates several key features of the interpretations, which is precisely why it makes such a convenient working example.

Before moving on, a rather compact, semi-formal derivation of $\DP$ in a Hilbert-style calculus is provided in Figure 1 - semi-formal because several of our inferences conflate a number of steps. Though we will refrain from carrying out any formal manipulations on proof trees, this will be used later to provide some insight into how the different techniques act on the proof in order to extract witnesses. Here $\forall$ax and $\forall$r refer to instances of the $\forall$-axiom and rule respectively (and similarly for the existential quantifier), while $\LEM$ denotes an instance of the law of excluded-middle and ctr an instance of contraction. The inference $(\ast)$ combines several steps which will typically not be relevant from a computational point of view.

\begin{figure}[t]
\label{fig-DP}
\begin{prooftree}
\AxiomC{$\neg P(k)\to P(k)\to P(m)$}
\RightLabel{$\forall$r}
\UnaryInfC{$\neg P(k)\to\forall m(P(k)\to P(m))$}
\RightLabel{$\exists$ax}
\UnaryInfC{$\neg P(k)\to\exists n\forall m(P(n)\to P(m))$}
\RightLabel{$\exists$r}
\UnaryInfC{$\exists k\neg P(k)\to\exists n\forall m(P(n)\to P(m))$}
\AxiomC{$P(m)\to P(0)\to P(m)$}
\RightLabel{$\forall$ax}
\UnaryInfC{$\forall k P(k)\to P(0)\to P(m)$}
\RightLabel{$\forall$r}
\UnaryInfC{$\forall k P(k)\to\forall m(P(0)\to P(m))$}
\RightLabel{$\exists$ax}
\UnaryInfC{$\forall k P(k)\to \exists n\forall m(P(n)\to P(m))$}
\RightLabel{$(\ast)$}
\BinaryInfC{$\exists k\neg P(k)\vee\forall k P(k)\to \exists n\forall m(P(n)\to P(m))\vee \exists n\forall m(P(n)\to P(m)) $}
\RightLabel{$\LEM$}
\UnaryInfC{$\exists n\forall m(P(n)\to P(m))\vee \exists n\forall m(P(n)\to P(m)) $}
\RightLabel{ctr}
\UnaryInfC{$\exists n\forall m(P(n)\to P(m))$}
\end{prooftree}
\caption{An informal Hilbert-style derivation of $\DP$}
\end{figure}

\section{Three famous interpretations}
\label{sec-interpretations}

In the first main part of the article, I give what is intended to be an accessible outline of three of the most famous computational interpretations of classical logic: Hilbert's epsilon calculus, Kreisel's modified realizability and G\"{o}del's Dialectica interpretation. The reader who wants to see a full exposition (and indeed full definitions) of these interpretations is encouraged to consult one of the many references I will provide on the way. My aim here is \emph{not} to give a comprehensive or detailed introduction, but to offer a case study which I hope will not only provide some insight into how the interpretations work in practise, but will hint at the deep connections between each of them.

It is important to point out that by focusing on just three interpretations, I inevitably exclude several significant approaches to program extraction, notably those originating from the French tradition such as Krivine's classical realizability \cite{Krivine(2009.0)}. This a reflection of the fact that I am somewhat familiar with the former, and much less so with the latter. Moreover, each of the interpretations I mention have a certain historical significance and are well-known outside of theoretical computer science, which I feel also justifies my choice. 

\subsection*{A note on precision}

Proof interpretations are formal maps on derivations, and as such, act in a very precise way on our proof of the drinkers paradox. Here, I give somewhat \emph{imprecise} versions of the formal extraction procedure, because I want to avoid superfluous bureaucratic details as much as possible. My emphasis here is on the salient features of the interpretations. A good example of this attitude is my cavalier approach to negative translations, which technically speaking have to be carefully specified and are, for example, sensitive to whether they are embeddings into intuitionistic or minimal logic. For the sake of clarity I skirt such issues here and only mention them in passing. Naturally, the reader curious about these subtleties will find a full explanation in the standard texts, and detailed references will be given here.

\subsection{The epsilon calculus}

Hilbert's epsilon calculus, developed in a series of lectures in the early 1920s, constitutes one of the first attempts at grounding classical first-order theories in a computational setting. Much like the Dialectica interpretation, it was first presented as a technique for proving the consistency of arithmetic, in which it comes hand-in-hand with the corresponding substitution method. For a more detailed introduction to the epsilon calculus itself, together with an extensive list of references, the reader is directed to \cite{AvZach(2002.0)}.

While the epsilon calculus remains a topic of interest in mathematical logic, philosophy and linguistics, concrete applications in mathematics or computer science remain limited in comparison to the interpretations that follow - a phenomenon which we discuss in Section \ref{sec-interlude}. Nevertheless, of the three interpretations we present in this section, the epsilon calculus is perhaps unrivalled in the simplicity and elegance of its main idea: To first bring predicate logic down to the propositional level by replacing all quantifier instances by a piece of syntax called an \emph{epsilon term}, and to then systematically eliminate these terms via an complex backtracking algorithm, thus leaving us in a finitary system in which existential statements are assigned explicit realizers.

Let's look at this idea more closely. Very informally, the epsilon calculus assigns to each predicate $A(x)$ an epsilon term $\epsilon x A(x)$, whose intended interpretation is a choice function which selects a witness for $\exists x A(x)$ whenever the latter is true. The idea is then to \emph{replace} the quantified formula $\exists x A(x)$ with $A(\epsilon x A(x))$, since under this interpretation these two formulas would be equivalent. The universal quantifier is dealt with similarly: bearing in mind that $\forall x A(x)\leftrightarrow \neg \exists x \neg A(x)$ over classical logic, we interpret $\forall x A(x)$ as $A(\epsilon x \neg A(x))$, since $\neg A(\epsilon x \neg A(x))$ holds only if $\neg \forall x A(x)$. 

We now want to transform proofs in predicate logic to proofs in the epsilon calculus. So what happens if we replace all instances of quantifiers in a proof with the corresponding epsilon term? It turns out that the quantifier rules are trivially eliminated with epsilon terms in place of quantifiers: If
\begin{prooftree}
\AxiomC{$\vdots$}
\noLine
\UnaryInfC{$A(x)\to B$}
\UnaryInfC{$\exists x A(x)\to B$}
\end{prooftree}
occurs in our proof then we simply replace the free variable $x$ with $\epsilon x A(x)$ and we get a derivation
\begin{prooftree}
\AxiomC{$\vdots$}
\noLine
\UnaryInfC{$A(\epsilon x A(x))\to B$.}
\end{prooftree}
On the other hand, if we have an instance of the quantifier \emph{axiom}
\begin{equation*}A(t)\to \exists x A(x)\end{equation*}
we now need to introduce a new axiom which governs the corresponding epsilon term, namely:
\begin{equation*}(\ast) \ \ \ A(t)\to A(\epsilon x A(x)).\end{equation*}
Axioms of the form $(\ast)$ are called \emph{critical axioms}. In simple terms, the epsilon calculus is the quantifier-free calculus we obtain by removing quantifiers and their axioms and rules, and adding epsilon terms together with the critical axioms. Note that for first-order theories like arithmetic things are already a bit more complicated, but for now we have enough structure to give an epsilon-style proof of the drinkers paradox! 

Referring to our derivation in Figure 1, let us use the following abbreviations:
\begin{equation*}
\begin{aligned}
\epsilon_k&:=\epsilon k \neg P(k) \\
\epsilon_m[n]&:=\epsilon m \neg (P(n)\to P(m)) \\
\epsilon_n&:=\epsilon n (P(n)\to P(\epsilon_m[n])) \end{aligned}
\end{equation*}
whose meaning in the epsilon calculus is indicated below:
\begin{equation*}
\begin{aligned}
P(\epsilon_k)&\leftrightarrow \exists k \neg P(k)\\
(P(n)\to P(\epsilon_m[n]))&\leftrightarrow \forall m(P(n)\to P(m))\\
P(\epsilon_n)\to P(\epsilon_m[\epsilon_n])&\leftrightarrow \exists n\forall m(P(n)\to P(m))
\end{aligned}
\end{equation*}
Note that the final formula is the drinkers paradox itself. 

Figure 2 shows the translation of our first-order derivation of $\DP$ into one in the epsilon calculus. The translation is quite straightforward, and uses nothing more than the principles sketched above: For each quantifier rule we simple replace the variable in question by the relevant epsilon term (and the rule vanishes), whereas each quantifier axiom is interpreted by the corresponding critical axiom, which is made explicit in Figure 2. The proof is shorter due to the elimination of the quantifier rules, and the instance of $\LEM$ now applies to the decidable predicate $P(\epsilon_k)$, and so is now computationally benign.

\begin{figure}[t]
\label{fig-eps}
\begin{prooftree}
\AxiomC{$\neg P(\epsilon_k)\to P(\epsilon_k)\to P(\epsilon_m[\epsilon_k])$}
\RightLabel{C2}
\UnaryInfC{$\neg P(\epsilon_k)\to (P(\epsilon_n)\to P(\epsilon_m[\epsilon_n]))$}
\AxiomC{$P(\epsilon_k)\to P(0)\to P(\epsilon_k)$}
\RightLabel{C1}
\UnaryInfC{$P(\epsilon_k)\to (P(0)\to P(\epsilon_m[0]))$}
\RightLabel{C3}
\UnaryInfC{$ P(\epsilon_k)\to (P(\epsilon_n)\to P(\epsilon_m[\epsilon_n]))$}
\BinaryInfC{$\neg P(\epsilon_k)\vee P(\epsilon_k)\to (P(\epsilon_n)\to P(\epsilon_m[\epsilon_n]))\vee (P(\epsilon_n)\to P(\epsilon_m[\epsilon_n])) $}
\RightLabel{$\LEM$}
\UnaryInfC{$(P(\epsilon_n)\to P(\epsilon_m[\epsilon_n]))\vee (P(\epsilon_n)\to P(\epsilon_m[\epsilon_n]))$}
\RightLabel{ctr}
\UnaryInfC{$P(\epsilon_n)\to P(\epsilon_m[\epsilon_n])$}
\end{prooftree}
Critical axioms:
\begin{enumerate}

\item[C1] $P(\epsilon_k)\to P(\epsilon_m[0])$

\item[C2] $(P(\epsilon_k)\to P(\epsilon_m[\epsilon_k]))\to (P(\epsilon_n)\to P(\epsilon_m[\epsilon_n]))$

\item[C3] $(P(0)\to P(\epsilon_m[0]))\to (P(\epsilon_n)\to P(\epsilon_m[\epsilon_n]))$

\end{enumerate}

\caption{An informal epsilon-style derivation of $\DP$}
\end{figure}
So far so good: We have transformed a proof in classical predicate logic to a quantifier-free version in which the main logical content is now encoded in the critical axioms. Can we get rid of these in some way in order to give the proof a computational interpretation? 

This is the role played by epsilon substitution, which forms the core of the epsilon calculus technique. In short: Any first-order proof can only use finitely many instances of the critical axioms. Therefore, if for all relevant formulas $A(x)$ which feature in the proof, we can find a \emph{concrete approximation} for the epsilon term $\epsilon x A(x)$, which rather than satisfying \emph{all} critical axioms $A(t)\to A(\epsilon x A(x))$, satisfy just those which are used in the proof, then we have a way of finding explicit witnesses for existential statements. The epsilon substitution method is an algorithm for eliminating critical formulas in this way.

It goes as follows: We first assign all relevant epsilon terms the canonical value $0$. We then examine our finite list of critical formulas until we find one which fails. This would mean that there is some term formula $A$ and term $t$ such that $A(t)$ is true but $A(\epsilon x A(x))$ false under our current assignment. But we can now use the this information to repair our epsilon term, by setting $\epsilon x A(x):=t$, which from now on serves as a witness for $\exists x A(x)$. The process is then repeated, until all critical formulas have been repaired.

This may sound quite straightforward, but epsilon substitution is in fact a complicated business! In the case of arithmetic, several erroneous algorithms were initially proposed, before a correct technique involving transfinite induction up to $\varepsilon_0$ was finally given by Ackermann in \cite{Ackermann(1940.0)} - a modern treatment of which is provided by Moser \cite{Moser(2006.0)}. The difficulty with the substitution method is primarily due to the presence of \emph{nested} epsilon terms, whereby trying to fix one critical axiom can suddenly invalidate several others which were previously considered fixed, leading to a subtle backtracking procedure. Further details of the general algorithm are far beyond the scope of this paper, but we \emph{are} able to provide a simple version for the case of the drinkers paradox!

First we should remark that in our proof, only two of our epsilon terms, $\epsilon_k$ and $\epsilon_n$, represent existential (as opposed to universal) quantifiers, so we only attempt to find approximations for these (we explain this in more detail below). Let us first try $\epsilon_k=\epsilon_n=0$. Then C2 and C3 are trivially satisfied, but not necessarily C1. In this case there are two possibilities: Either we get lucky and $P(0)\to P(\epsilon_m[0])$ holds, in which case we're done, or our guess fails because $P(0)\wedge \neg P(\epsilon_m[0])$. We now learn from our failure and repair the broken epsilon terms, setting $\epsilon_k:=\epsilon_m[0]$. Now C1 holds and C3 remains the same as before, but under our new assignment C2 becomes
\begin{equation*}
(P(\epsilon_m[0])\to P(\epsilon_m[\epsilon_m[0]]))\to (P(0)\to P(\epsilon_m[0]))
\end{equation*}
which from our assumption $P(0)\wedge \neg P(\epsilon_m[0])$ is now false. So we repair this in turn and set $\epsilon_n:=\epsilon_m[0]$. A quick run through of each critical axiom under the assignment $\epsilon_k=\epsilon_n:=\epsilon_m[0]$ reveals that we're done.
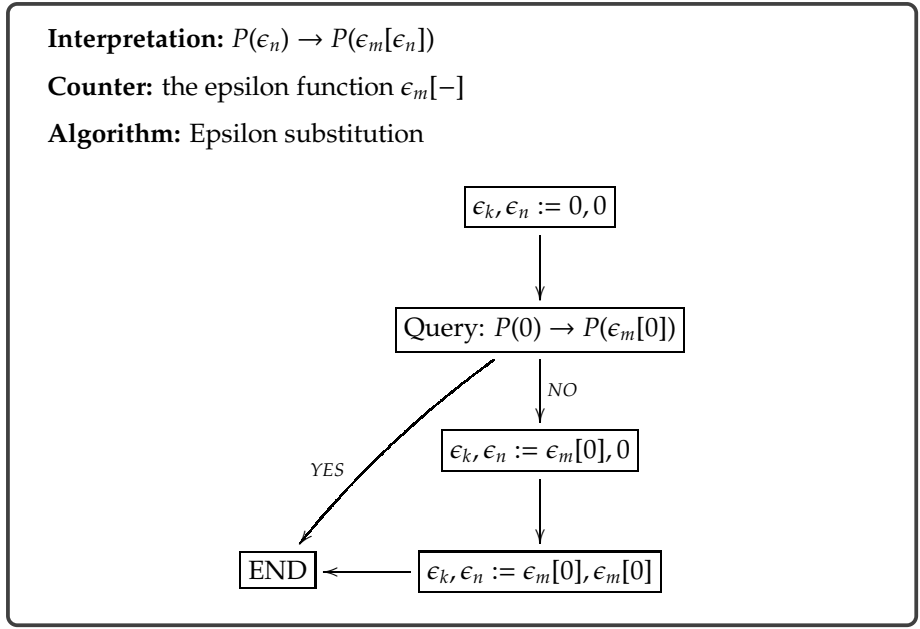
\begin{figure}[t]
\begin{tcolorbox}\textbf{Interpretation:} $P(\epsilon_n)\to P(\epsilon_m[\epsilon_n])$\medskip

\textbf{Counter:} the epsilon function $\epsilon_m[-]$\medskip

\textbf{Algorithm:} Epsilon substitution
\[\xymatrix{& \fbox{$\epsilon_k,\epsilon_n:=0,0$}\ar[d] \\  & \fbox{Query: $ P(0)\to P(\epsilon_m[0])$}\ar@/_/[ldd]_>>>>>>>>>{YES}\ar[d]^{NO}\\  & \fbox{$\epsilon_k,\epsilon_n:=\epsilon_m[0],0$}\ar[d] \\ \fbox{END} & \fbox{$\epsilon_k,\epsilon_n:=\epsilon_m[0],\epsilon_m[0]$}\ar[l] }\]

\end{tcolorbox}
\caption{Interpretation of $\DP$ via epsilon calculus}
\end{figure}

The substitution method generalises this strategy of guessing, learning from the failure of our guesses, and repairing. As such, despite its complexity, it offers a beautifully clear computational semantics for classical logic: The building of approximations to non-computational objects via a game of trial and error. This has inspired a number of more modern approaches to program extraction, which we will mention later.


As we already pointed out, there are no critical formulas for $\epsilon_m[n]$. Rather, this plays the role of a function variable, and our witness for $\epsilon_n$ is constructed relative to this variable. Without wanting to go into more detail here, $\epsilon_m[n]$ is a placeholder which represents how the drinkers paradox might be used as a lemma in the proof of another theorem. As such, we do not have an ideal witness for $\exists n$ in the drinkers paradox, but rather an approximate witness relative to $\epsilon_m[n]$. We will see this phenomenon repeat itself throughout this paper, where in each setting there will be a specific structure which plays the role of a `counter argument', which represents the universal quantifier and against which our approximate witness is computed.  

We summarise our epsilon-style interpretation of $\DP$ in Figure 3, giving a flowchart representation of the corresponding substitution algorithm.\smallskip

\subsection{Modified realizability}
\label{sec-interpretations-mr}

Before we move on to the world of realizability, let us consider the following elegant interpretation of $\DP$ as a game between two players, who are traditionally named $\exists$loise and $\forall$belard. $\exists$loise's goal is to establish the truth of $\DP$ by choosing a witness for $\exists n$, whereas her opponent $\forall$belard is tasked with disproving $\DP$ by producing a counter witness for $\forall m$. Fortunately for $\exists$loise, the rules of the game dictate that she can backtrack and change her mind!

So how might the game go? $\exists$loise begins by picking an arbitrary witness for $\exists n$, let's say $n:=0$, thereby claiming that $\forall m(P(0)\to P(m))$ is true. $\forall$belard responds by playing some $m$ with the aim of showing that $\exists$loise's guess was wrong. There are now two possibilities: Either $P(0)\to P(m)$ is true and $\exists$loise was right all along, in which case she wins, or $\forall$abelard's challenge was successful and $P(0)\wedge\neg P(m)$ holds. But now $\exists$loise responds by simply changing her mind and playing $n:=m$. Now any further play $m'$ from $\forall$belard is destined to fail since $P(m)\to P(m')$ will always be true! 

What we have done is describe a winning strategy for $\exists$loise in a quantifier-game corresponding to $\DP$. That there is a correspondence between classical validity and winning strategies in this sense is well-known and widely researched, dating back to e.g. Novikoff in \cite{Novikoff(1943.0)}, and today game semantics is an important topic in logic. Here it will form a useful sub-theme and will help us characterise the behaviour of certain functionals which arise from classical proofs. 

Modified realizability is one of several forms of realizability which arose as a concrete implementation of the Brouwer-Heyting-Kolmgorov (BHK) interpretation of intuitionistic logic. Introduced by Kreisel in \cite{Kreisel(1959.0)}, modified realizability works in a \emph{typed} setting, and allows us to transform proofs in intuitionistic arithmetic to terms in the typed lambda calculus.

As a clean and elegant formulation of the BHK interpretation in the typed setting, variants of modified realizability have proven extremely popular techniques for extracting programs from proofs. In particular, refinements of the interpretation form the theoretical basis for the Minlog system \cite{Minlog}, a proof assistant which automates program extraction and is primarily motivated by the synthesis of verified programs (see e.g. \cite{BergLFS(2015.0),BergMiySch(2016.0)} for examples of this). For a comprehensive account of the interpretation itself, the reader is directed to e.g. \cite[Chapter 5]{Kohlenbach(2008.0)} or \cite[Chapter 7]{SchWai(2011.0)}.

Similarly to the epsilon calculus, modified realizability consists of two components: An interpretation and a soundness proof. The interpretation maps each formula $A$ in Heyting arithmetic to a new formula $\mr{x}{A}$, where $x$ is a potentially empty tuple of variables whose length and type depends on the structure of $A$. The interpretation is quite simple, so we state it in Figure 4. 
\begin{figure}[h]
\begin{equation*}
\begin{aligned}
\mr{x}{A}&:\equiv A\mbox{ for prime formulas $A$}\\
\mr{x,y}{A\wedge B}&:\equiv \mr{x}{A}\wedge \mr{y}{B}\\
\mr{b,x,y,}{A\vee B}&:\equiv (b=0\to \mr{x}{A})\wedge (b\neq 0\to\mr{y}{B})\\
\mr{f}{A\to B}&:\equiv \forall x(\mr{x}{A}\to \mr{fx}{B})\\
\mr{z,y}{\exists x A(x)}&\equiv \mr{y}{A(z)}\\
\mr{f}{\forall x A(x)}&\equiv \forall x(\mr{fx}{A(x)})
\end{aligned}
\end{equation*}
\caption{Modified realizability}
\end{figure}

The soundness proof for modified realizability states that whenever $A$ is provable in Heyting arithmetic, then there is some term $t$ of System T (i.e. the typed lambda calculus of primitive recursive functionals in all finite types) satisfying $\mr{t}{A}$, which can moreover be formally extracted from the proof:
\begin{equation*}
\HA\vdash A\mbox{ \ \  implies \ \  } \mbox{System T}\vdash \mr{t}{A}
\end{equation*}
So far, what we have described applies only to intuitionistic theories, so we need an additional step to extend the interpretation to classical logic. This turns out to be rather subtle, and usually relies on a combination of the G\"{o}del-Getzen negative translation together with some variant of the so-called Dragalin/Friedman/Leivant trick, also known as the A-translation. A detailed discussion of this technique is beyond the scope of this paper, but we briefly present a variant due to \cite{BergSch(1995.0)} and describe how it acts on $\DP$. 

First the negative translation. Negative translations are well-known methods for embedding classical logic in intuitionistic logic. There are several variants of the translation, each of which involves strategically inserting double negations in certain places in a logical formula in such a way that if $A$ is provable in classical logic, then $A^N$ is provable intuitionistically. In particular, we would have
\begin{equation*}\PA\vdash A\mbox{ \ \  implies \ \  } \HA\vdash A^N\end{equation*}
Negative translations are widely used and and the relationship between the many varieties is well understood (see \cite{FerGOli(2012.0)} for a detailed study). We deliberately avoid going into more detail and specifying a translation to use here. We simply state without justification that a suitable negative translation of the drinkers paradox is the following:
\begin{equation*}
\DP^N:\equiv \neg\neg\exists n\forall m(P(n)\to \neg\neg P(m))
\end{equation*}
which is provable in intuitionistic (and in fact minimal) logic. Now to the A-translation. The reason we need an intermediate step in addition to the negative translation is that the negative translation alone results in a formula with no computational meaning at all: Since $\bot$ is a prime formula, we have
\begin{equation*}
\mr{f}{\neg A}\equiv \forall x(\mr{x}{A}\to \bot)
\end{equation*}
i.e. $f$ is just an empty tuple. In order to make realizability `sensitive' to the negative translation, we treat $\bot$ as a new predicate, which in particular has a special realizability interpretation
\begin{equation*}
\mr{x}{\bot}
\end{equation*}
where $x$ has some predetermined type $\tau$. With this adjustment we would have
\begin{equation*}
\mr{f}{\neg A}\equiv \forall x(\mr{x}{A}\to \mr{fx}{\bot})
\end{equation*}
where $fx$ has the non-empty type $\tau$. Now that realizability interacts with negated formulas, we are ready to go! The first step is to examine the negated principle $\DP^N$, and carefully apply the clauses of Figure 4 together with the special interpretation of $\bot$.

We now do this in detail. We first assume that the decidable predicate $P(n)$ can be coded as a prime formula $t(n)=0$ and thus has an empty realizer. So starting with the inner formula it is not to difficult to work out that
\begin{equation*}
\begin{aligned}
\mr{f}{(P(n)\to \neg\neg P(m))}&\equiv P(n)\to \forall a((P(m)\to \mr{a}{\bot})\to \mr{fa}{\bot})
\end{aligned}
\end{equation*}
where $f:\tau\to\tau$. The next step is to treat the quantifiers, and we end up with
\begin{equation*}
\begin{aligned}
&\mr{e,g}{\exists n\forall m (P(n)\to \neg\neg P(m))}\\
&\equiv \forall m(\mr{gm}{(P(e)\to \neg\neg P(m))})\\
&\equiv \underbrace{\forall m (P(e)\to \forall a((P(m)\to \mr{a}{\bot})\to \mr{gma}{\bot}))}_{Q(e,g)}
\end{aligned}
\end{equation*}
where for convenience we label this formula $Q(e,g)$ as indicated. Note that types of these realizers are given by $e:\NN$ and $g:\NN\to\tau\to\tau$. Finally, interpreting the whole formula yields
\begin{equation}
\label{mr-main}\begin{aligned}
&\mr{\Phi}{\neg\neg\exists n\forall m(P(n)\to \neg\neg P(m))}\\
&\equiv \forall p(\mr{p}{\neg\exists n\forall m (P(n)\to \neg\neg P(m))}\to \mr{\Phi p}{\bot})\\
&\equiv \forall p(\forall e,g(Q(e,g)\to\mr{peg}{\bot})\to \mr{\Phi p}{\bot}) 
\end{aligned}
\end{equation}
where now $p:\sigma$ and $\Phi:\sigma\to\tau$ for $\sigma:=\NN\to (\NN\to\tau\to\tau)\to\tau$.

\begin{figure}
\begin{tcolorbox}\textbf{Interpretation:} $\mr{\Phi}{\neg\neg \exists n\forall m(P(n)\to\neg\neg P(m))}$\medskip

\textbf{Counter:} $\forall$belard i.e. $\lambda m,a$ in $\Phi$ \medskip

\textbf{Algorithm:} Unwinding $\mr{\Phi p}{\bot}$
\[\xymatrix{& \fbox{$\mr{p0h_p}{\bot}$}\ar[d]^{\mbox{\scriptsize $\exists$loise plays $e,g:=0,h_p$}} \\ & \fbox{$Q(0,h_p)$}\ar[d]^{\mbox{\scriptsize $\forall$belard plays $m,a$}} \\ & \fbox{$P(0)\to (P(m)\to \mr{a}{\bot})\to \mr{h_pm0}{\bot}$}\ar[d] \\ & \fbox{Query: $P(0)\to P(m)$}\ar@/_2pc/[lddd]_>>>>>>>>>>>>{YES}\ar[d]^{NO} \\ & \fbox{$\mr{pm(\lambda n,b.b)}{\bot}$}\ar[d]^{\mbox{\scriptsize $\exists$loise plays $e,g:=m,\lambda n,b.b$}} \\ & \fbox{$Q(m,\lambda n,b.b)$}\ar[d]^{\mbox{\scriptsize $\forall$belard plays $n,b$}} \\ \fbox{END} & \fbox{$P(m)\to (P(n)\to \mr{b}{\bot})\to \mr{b}{\bot}$}\ar[l]}\]

\end{tcolorbox}
\caption{Modified realizability interpretation of $\DP$}
\end{figure}
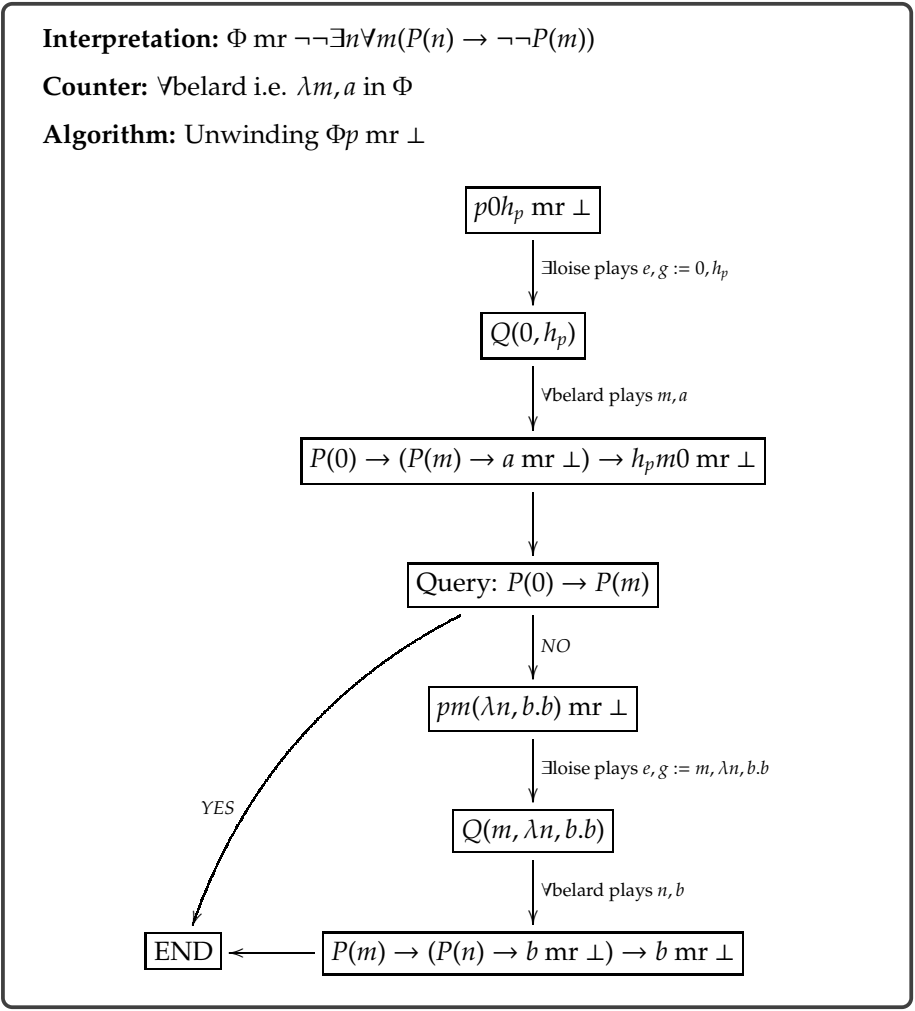

Just like we did not give a formal description of the epsilon substitution method, we will not present in detail how a concrete functional $\Phi$ satisfying the above is rigorously extracted from the proof of the negative translation of $\DP$. Rather, we simply present a term which does the trick and carefully explain why. Let's define
\begin{equation*}\Phi p:=p0h_p\mbox{ \ \ \ where \ \ \ }h_p:=\lambda m^\NN,a^\tau\; \left(\mbox{$a$ if $P(0)\to P(m)$ else $pm(\lambda n^\NN,b^\tau.b)$}\right).\end{equation*}
We need to prove that $\Phi$ satisfies $(\ref{mr-main})$ above. In order to do this, we have to simultaneously unwind $(\ref{mr-main})$ together with the definition of $\Phi$. Our goal is to show that $\mr{\Phi p}{\bot}$ whenever $p$ satisfies the premise of $(\ref{mr-main})$. So let's assume the latter, i.e.
\begin{equation}\label{mr-prem}\forall e,g(Q(e,g)\to \mr{peg}{\bot})\end{equation}
and first instantiate $e,g:=0,h_p$, which gives us
\begin{equation*}Q(0,h_p)\to \mr{\underbrace{p0h_p}_{=\Phi p}}{\bot}\end{equation*}
It is now enough to show that $Q(0,h_p)$ holds, since then we have $\mr{\Phi p}{\bot}$ as indicated. Referring back to the definition of $Q(e,g)$ above, this in turn boils down to showing that, for any $m,a$:
\begin{equation}\label{mr-aux} P(0)\to (P(m)\to \mr{a}{\bot})\to \mr{h_p ma}{\bot}.\end{equation}
There are now two cases. In the first, $P(0)\to P(m)$ holds, which means that $h_pma=a$ and so $(\ref{mr-aux})$ becomes
\begin{equation*}P(0)\to (P(m)\to\mr{a}{\bot})\to \mr{a}{\bot}.\end{equation*}
This is now true, since either $P(0)$ is false and it follows trivially or $P(m)$ holds, and so $\mr{a}{\bot}$ follows from the assumption $P(m)\to\mr{a}{\bot}$. 

In the second case we have $P(0)\wedge\neg P(m)$, which means that $h_pma=pm(\lambda n,b.n)$ and $(\ref{mr-aux})$ becomes
\begin{equation*}
P(0)\to (P(m)\to \mr{a}{\bot})\to \mr{pm(\lambda n,b.b)}{\bot},
\end{equation*}
which appears less simple to validate. But the trick is to now go back to our main assumption (\ref{mr-prem}), this time instantiating $e,g:=m,\lambda n,b.n$. We are now done so long as $Q(m,\lambda n,b.b)$ holds, which written out properly is
\begin{equation*}P(m)\to (P(n)\to \mr{b}{\bot})\to \mr{b}{\bot}.\end{equation*}
But this now trivially true for any $n,b$ since by assumption we have $\neg P(m)$! Therefore climbing back up: We have established $\mr{pm(\lambda n,b.b)}{\bot}$ and hence $\mr{h_pma}{\bot}$, which in turn proves $Q(0,h_p)$ and thus $\mr{p0h_p}{\bot}$.

At first glance, the verification of our realizer looks somewhat convoluted, jumping back and forth between our various hypothesis until one by one they have been discharged. However, things become much clearer if we visualise the above argument in terms of the game sketched at the beginning of the section. 

Let's look at it again, this time giving each step a game-theoretic reading. In order to verify $\mr{\Phi p}{\bot}$ we set $e,g:=0,h_p$ in $(\ref{mr-prem})$, which can be seen as a first attempt by $\exists$loise to prove $\DP$. In this context, $\forall$belard's job is to disprove $Q(0,h_p)$, which he does by choosing some $m,a$ and hoping that $(\ref{mr-aux})$ fails. Either $P(0)\to P(m)$ is true, in which case $(\ref{mr-aux})$ holds and $\exists$loise wins, or $\forall$belard chose a good counterexample and $P(0)\wedge\neg P(m)$. $\exists$loise now appeals to the premise of $(\ref{mr-prem})$ a second time, setting $e,g:=m,\lambda n,b.n$, and wins unless $Q(m,\lambda n,b.b)$ can be shown false. But under our assumption $\neg P(m)$ this is impossible, and so $\exists$loise has a winning strategy. 

What is interesting here is how the winning strategy is encoded by the components of our term $\Phi$: $\exists$loise's moves are arguments for the term $p$, whereas $\forall$belard's move is represented by the internal function abstraction $\lambda m,a$. 

The reader interested in exploring the the connection between realizability, negative translations and games could take \cite{Coquand(1995.0)} as a starting point. A fascinating illustration of this phenomenon in the case of the axiom of choice is provided by \cite{BBC(1998.0)}, which also inspired my own work in \cite{Powell(2015.0)}. 

We finish off this section, as before, with a diagrammatic representation of the algorithm which underlies this realizability interpretation of $\DP$. Note that where for the epsilon calculus the epsilon function $\epsilon_m[-]$ played the role of the `counter', here the same thing is represented by the player $\forall$belard, or more precisely, the function abstraction within our term $\Phi$. 

\subsection{Dialectica interpretation}
\label{sec-interpretations-dial}

We now come to our final computational interpretation of classical logic: G\"{o}del's functional, or Dialectica interpretation. The name `Dialectica' refers to the journal in which the interpretation was first published in 1958 \cite{Goedel(1958.0)}, though the interpretation had been conceived as early as the 1930s, and had been presented in lectures from the early 1940s. An English translation of the original article can be found in Volume II of the \emph{Collected Works} \cite{Goedel(1990.0)}, which is preceded by an illuminating introduction by Troelstra.

The original purpose of the Dialectica interpretation was to produce a relative consistency proof for Peano arithmetic. The interpretation maps the first-order theory of arithmetic to the primitive recursive functionals in finite types, thereby showing that the consistency of the former follows from that of the latter. The interpretation was soon extended to full classical analysis by Spector, in another groundbreaking article \cite{Spector(1962.0)}, which is also a fascinating read due to the extensive footnotes from Kreisel, who put together and completed the paper following Spector's early death in 1961.

Much like the epsilon calculus, then, the Dialectica interpretation has its origins in Hilbert's program and the problem of consistency, though in contrast it has achieved great success as a tool in modern applied proof theory, being central to the proof mining program initiated by Kreisel in the 1960s and brought to maturity by Kohlenbach from the 1990s. 

For a comprehensive introduction to the interpretation the reader is directed to the Avigad and Feferman's \emph{Handbook} chapter \cite{AvFef(1998.0)} or the textbook of Kohlenbach \cite{Kohlenbach(2008.0)}, particularly Chapters 8-10. The latter is also the standard reference for \emph{applications} of the Dialectica interpretation in mathematics (and applied proof theory in general). 

The basic set up of the interpretation bears a close resemblance to modified realizability, although there are crucial differences between the two. The Dialectica assigns to each formula $A$ of Heyting arithmetic a new formula of the form $\exists x\forall y\dt{A}{x}{y}$, where $\dt{A}{x}{y}$ is quantifier-free and $x$ and $y$ are tuples of variables whose length and type depend on the structure of $A$. The inner part of the interpretation is defined by induction on formulas, which we give in Figure 6 below.

\begin{figure}[h]
\begin{equation*}
\begin{aligned}
\dt{A}{x}{y}&:\equiv A\mbox{ for prime formulas $A$}\\
\dt{A\wedge B}{x,u}{y,v}&:\equiv \dt{A}{x}{y}\wedge \dt{B}{u}{v}\\
\dt{A\vee B}{b^\NN,x,u}{y,v}&:\equiv (b=0\to \dt{A}{x}{y})\wedge (b\neq 0\to\dt{B}{u}{v})\\
\dt{A\to B}{f,g}{x,v}&:\equiv \dt{A}{x}{gxv}\to\dt{B}{fx}{v}\\
\dt{\exists z A(z)}{z,x}{y}&:\equiv \dt{A(z)}{x}{y}\\
\dt{\forall z A(z)}{f}{z,y}&:\equiv \dt{A(z)}{fz}{y}
\end{aligned}
\end{equation*}
\caption{The Dialectica interpretation}
\end{figure}

When one first sees the definition of the Dialectica interpretation, it is immediate that it departs from the usual BHK style on which realizability is based, and the interpretation of implication in particular seems rather ad-hoc until one understands where it comes from! The point here is that Dialectica interpretation acts as a kind of Skomelisation, pulling all the quantifiers to the front of the formula but preserving logical validity, so that
\begin{equation*}A\leftrightarrow \exists x\forall y\dt{A}{x}{y}\end{equation*}
provably in the usual higher-type formulation of classical logic plus a quantifier-free form of choice. 

For most of the logical connectives, there is one obvious way of defining the interpretation. However, in the case of implication we have several ways of pulling the quantifiers out to the front, which make use of classical logic to a greater or lesser degree. G\"{o}del chose the option which Skolemises implication in the `least non-constructive way', using only Markov's principle and a weak independence of premise. A detailed explanation of all this is given in \cite[Section 2.3]{AvFef(1998.0)} and \cite[pp. 127--129]{Kohlenbach(2008.0)}. I highlight it here because the treatment of implication can often seem mystifying until one sees that it is defined the way it is for a reason!

As with modified realizability, the Dialectica interpretation comes equipped with a soundness proof: If $A$ is provable in Heyting arithmetic then there is some term $t$ of System T satisfying $\forall y\dt{A}{t}{y}$, and moreover, this term can be extracted from the proof of $A$:
\begin{equation*}
\HA\vdash A\mbox{ \ \  implies \ \  } \mbox{System T}\vdash \forall y\dt{A}{t}{y}.
\end{equation*}
We are now in a similar situation to the last section: We have a computational interpretation of intuitionistic arithmetic, which we now need to extend in some way to deal with classical logic. It turns out that the same trick works: We just combine the Dialectica interpretation with the negative translation. But crucially, for the Dialectica we do not need an intermediate step corresponding to the A-translation, since negated formulas are already considered computational thanks to the interpretation of implication: We have
\begin{equation}
\label{dial-neg}
\neg A\mbox{ is interpreted as } \exists g\forall x\dt{A\to \bot}{g}{x}\equiv \exists g\forall x(\dt{A}{x}{gx}\to \bot)
\end{equation}
On top of this, the Dialectica allows us to simplify our negative translated formulas considerably. If $A$ is a prime formula then
\begin{equation*}
\neg\neg A\mbox{ is interpreted as } \neg\neg A\mbox{ which is intuitionistically equivalent to } A
\end{equation*}
Therefore when interpreting a complicated formula we can continually remove inner double negations which apply only to quantifier-free inner parts, which saves us a lot of effort!

The reader may wonder why we didn't go ahead and do this in the case of modified realizability. This is another consequence of not needing to treat $\bot$ as a special symbol: The variant of the $A$ translation we gave uses implicitly that $\DP^N$ is provable in \emph{minimal} logic, in which case we can replace $\bot$ with anything and the proof still goes through. However, the removal of double negations before quantifier-free formulas requires ex-falso quodlibet and hence the simplified form of $\DP^N$ we will consider below cannot be dealt with by the $A$ translation. For more on this rather subtle point see \cite[Chapter 14]{Kohlenbach(2008.0)}.

So let's now focus on how the Dialectica interpretation deals with $\DP$, or more specifically, its negative translation, we which already gave in the previous section as
\begin{equation*}\neg\neg \exists n\forall m (P(n)\to \neg\neg P(m)).\end{equation*}
As discussed above, the first remarkable difference with realizability is that the inner double negation essentially vanishes. Since $P(n)\to \neg\neg P(m)$ is quantifier-free, it can be encoded as a prime formula and we have
\begin{equation*}(P(n)\to \neg\neg P(m))\mbox{ is interpreted as } (P(n)\to \neg\neg P(m))\leftrightarrow (P(n)\to P(m))\end{equation*}
since in Heyting arithmetic $\neg\neg P(m)\leftrightarrow P(m)$. Therefore, looking at how the Dialectica interprets quantifiers and removing our inner double negation in this way, we have that
\begin{equation*}\exists n\forall m(P(n)\to \neg\neg P(m))\mbox{ is interpreted as }\exists n\forall m (P(n)\to P(m))\end{equation*}
Therefore in order to deal with $\DP^N$, we need to interpret a formula of the form $\neg\neg B$ where $B:\equiv \exists n\forall m(P(n)\to P(m))$ is an $\exists\forall$ formula. It's here that the interpretation of implication comes into play.
To interpret the outer negations, let's first look at $\neg B$. Following (\ref{dial-neg}), this is interpreted as
\begin{equation*}\begin{aligned}
\exists g\forall n\neg (P(n)\to P(gn))
\end{aligned}\end{equation*}
and therefore negating a second time, the interpretation of $\neg\neg B$ is given by
\begin{equation*}
\exists \Phi\forall g \neg\neg (P(\Phi g)\to P(g(\Phi g)))\leftrightarrow \exists \Phi\forall g (P(\Phi g)\to P(g(\Phi g)))
\end{equation*}
Putting this all together, we have that the Dialectica interpretation of the negative translation of $\DP$ is, modulo the elimination of inessential double negations, the following:
\begin{equation}
\label{dial-int}
\exists \Phi\forall g(P(\Phi g)\to P(g(\Phi g)).
\end{equation}
Our challenge is to find a functional $\Phi$ which satisfies (\ref{dial-int}). It turns out that such a functional is a lot easier to write down than that for modified realizability, and so this time we want to hint - very informally nevertheless - at how such a term can be extracted from the semi-formal proof given in Figure 1, as it highlights some important features of the \emph{soundness} of the Dialectica interpretation.

The main idea is to extract our functional $\Phi$ recursively over the structure of the proof. Roughly speaking this goes as follows: There are two main branches of the proof, which are then combined using excluded-middle to yield two potential witnesses, which are reduced to an exact witness via contraction, which computationally speaking induces a case distinction. 

Let's first focus on the left-hand branch, which establishes
\begin{equation*}\exists k\neg P(k)\to \exists n\forall m(P(n)\to P(m)).\end{equation*}
The functional interpretation of the negative translation of the above is equivalent to
\begin{equation*}\exists \Psi_1\forall k,g(\neg P(k)\to P(\Psi_1 kg)\to P(g(\Psi_1 kg)))\end{equation*}
which is very easily satisfied by $\Psi_1 kg:=k$ i.e.
\begin{equation*}
\forall k,g(\neg P(k)\to P(k)\to P(gk))
\end{equation*}
Now we turn to the right-hand branch, which establishes
\begin{equation*}\forall k P(k)\to \exists n\forall m (P(n)\to P(m))\end{equation*}
\begin{figure}[t]
\begin{prooftree}
\AxiomC{$\neg P(k)\to P(k)\to P(m)$}
\RightLabel{$\forall$r}
\UnaryInfC{$\forall m(\neg P(k)\to P(k)\to P(m))$}
\RightLabel{$\exists$ax}
\UnaryInfC{$\forall g(\neg P(k)\to P(k)\to P(gk))$}
\RightLabel{$\exists$r}
\UnaryInfC{$\forall k,g(\neg P(k)\to P(k)\to P(gk))$}
\AxiomC{$P(m)\to P(0)\to P(m)$}
\RightLabel{$\forall$ax}
\UnaryInfC{$P(m)\to P(0)\to P(m)$}
\RightLabel{$\forall$r}
\UnaryInfC{$\forall m( P(m)\to P(0)\to P(m))$}
\RightLabel{$\exists$ax}
\UnaryInfC{$\forall g( P(g0)\to P(0)\to P(g0))$}
\BinaryInfC{$\forall g(\neg P(g0)\vee P(g0)\to (P(g0)\to P(g(g0)))\vee (P(0)\to P(g0))) $}
\RightLabel{$\LEM$}
\UnaryInfC{$\forall g((P(g0)\to P(g(g0)))\vee (P(0)\to P(g0)))$}
\RightLabel{ctr}
\UnaryInfC{$\forall g(P(\Phi g)\to P(g(\Phi g)))$}
\end{prooftree}
\caption{An informal extraction of $\Phi$}
\end{figure}
This is interpreted as
\begin{equation*}\exists \Psi_2,\Psi_3\forall g(P(\Psi_3 g)\to P(\Psi_2 g)\to P(g(\Psi_2 g))) \end{equation*}
which is also easily satisfied by $\Psi_2 g=0$ and $\Psi_3 g=g0$ i.e.
\begin{equation*}
\forall g(P(g0)\to P(0)\to P(g0))
\end{equation*}
We now mimic the combining of the two branches by setting $k:=g0$ (note that this would formally correspond to several steps). We obtain
\begin{equation*}
\forall g(\neg P(g0)\vee P(g0) \to (P(g0)\to P(g(g0)))\vee (P(0)\to P(g0)))
\end{equation*}
and so eliminating the now quantifier-free instance of LEM we end up with
\begin{equation*}
\forall g((P(g0)\to P(g(g0)))\vee (P(0)\to P(g0))).
\end{equation*}
The final instance of contraction is dealt with by carrying out a definition-by-cases: Setting
\begin{equation*}\Phi g:=\mbox{$0$ if $P(0)\to P(g0)$ else $g0$}\end{equation*}
we have
\begin{equation*}
\forall g(P(\Phi g)\to P(g(\Phi g)))
\end{equation*}
and so we're done. Of course, with a little thought we could have come up with such a program without going through the formal extraction. However, we want to illustrate how the construction of realizers mimics the formal proof, which we sketch via our proof tree in Figure 7. Note that, technically speaking, the application of the negative translation to the proof in Figure 1 would result in a new, bigger proof tree which establishes $\DP^N$, and over which our realizer would be extracted. However, for readability we conflate this heavily in Figure 7, since it's only the main structure we want to highlight.

Figure 8 gives our usual summary of the algorithm underlying the interpretation. In this case it is quite succinct: Our functional $\Phi$ is nothing more than a straightforward case distinction and our `counter' is just a simple function $g$. In this way, our interpretation is neither embellished by a nested function calls as in modified realizability, or with an explicit backtracking algorithm as for the epsilon calculus. 

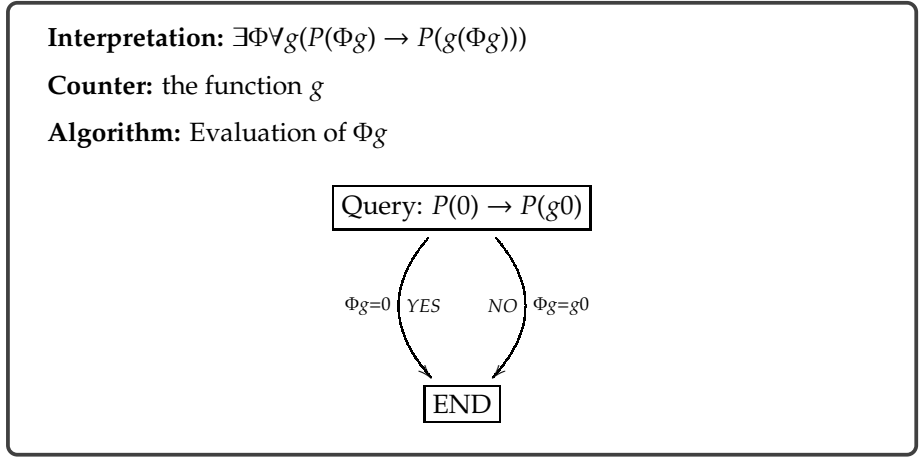
\begin{figure}[h]
\begin{tcolorbox}\textbf{Interpretation:} $\exists \Phi\forall g(P(\Phi g)\to P(g(\Phi g)))$\medskip

\textbf{Counter:} the function $g$\medskip

\textbf{Algorithm:} Evaluation of $\Phi g$
\[\xymatrix{\fbox{Query: $P(0)\to P(g0)$}\ar@/_2pc/[dd]^{YES}_{\Phi g=0}\ar@/^2pc/[dd]_{NO}^{\Phi g=g0} \\ \\ \fbox{END} }\]

\end{tcolorbox}
\caption{Dialectica interpretation of $\DP$}
\end{figure}

\subsection{Final remarks}

It will hopefully not have escaped the reader that in the simple case of the drinkers paradox, all three interpretations arrive at essentially the same basic algorithm:
\begin{itemize}

\item Try a default value $n:=0$;

\item Query the truth value of $P(0)\to P(m)$ where $m$ is some counter value obtained from the environment;

\item In the case of failure, update $n:=m$.

\end{itemize}

This useful feature of the drinkers paradox allows us to focus on the very different ways in which each interpretation describes the underlying algorithm. Notably, the interaction with a counter value is present in each of the above examples, but encoded using a variety of structures, from a simple function in the case of the Dialectica to an internal abstraction in the case of modified realizability. At this point, it is perhaps worth pausing a moment to discuss these differences and the way in which they have affected how each interpretation has been used since Hilbert's program.

\section{Interlude: A general perspective on proof interpretations}
\label{sec-interlude}

It may seem remarkable that as we slowly approach a century since the emergence of the epsilon calculus, a comprehensive comparison and assessment of the well-established computational interpretations of classical logic is still lacking - though a number of studies in this direction (such as \cite{Trifonov(2011.0)}) have been produced. 

Comparing proof interpretations is difficult. For a start, they are acutely sensitive to the way in which they are applied: An unnecessary double negation that would be routinely removed by a human applying a proof interpretation by hand could result in a considerable blow-up in complexity for a program formally extracted by a machine. Thus the Dialectica interpretation as a tool in proof mining is different to the Dialectica interpretation as an extraction mechanism in a proof assistant. Closely related to this is the fact that proof interpretations are in a constant state of flux, undergoing adaptations and refinements in quite specific directions as they become established as tools in contrasting areas of application. So when comparing, for instance, the Dialectica to modified realizability, we have to quite carefully specify which of the many variants we have in mind.

Another problem lies in deciding the \emph{nature} of the comparison we wish to make. For example, it is clear that in terms of their definitions, modified realizability and dialectica belong very much to the same family of interpretations (an impression which is made precise in \cite{Oliva(2006.0)}), whereas the epsilon calculus is different species altogether. Nevertheless, in terms of the programs they produce for $\DP$, it can be argued that the Dialectica has more in common with the epsilon calculus than it does with realizability. Without wanting to read too much into our simple case study, the point is simply that there is more to a proof interpretation than meets the eye, and a similarity in the definitional structure of two interpretations is not necessarily reflected when one examines the programs they produce.

As such, it is difficult to make sweeping claims about the relationship between proof interpretations without a certain amount of qualification. Nevertheless, it is undoubtedly the case that over the years, each of the three interpretations of classical logic mentioned above has developed a reputation, closely tied to the domains in which they primarily feature, and moreover posses certain characteristics which make them appealing for certain applications, and less appealing for others.

Perhaps the most striking feature of Hilbert's epsilon calculus is that it contains, implicitly, an elegant computational explanation of classical reasoning. Classical logic allows us to construct `ideal objects' which are represented by epsilon terms, and the substitution methods shows us how to build approximations to these objects via a form of trial and error. In this way, the epsilon calculus can be seen as a direct precursor to much later work such as Coquand's game semantics \cite{Coquand(1995.0)}, Avigad's update procedures \cite{Avigad(2002.0)}, the Aschieri/Berardi learning-based realizability \cite{AscBer(2010.0)} and much more besides, which are all concerned in some way with a computational semantics of classical arithmetic via backtracking, or learning, a topic which we explore in more detail later. 

All of this aside, the epsilon calculus plays active role in structural proof theory, where it has deep links with cut elimination \cite{Mints(2008.0)}, and the variants of the substitution method are still studied, particularly with regards to complexity issues \cite{MosZac(2006.0)}. More recently, the syntactic representation of quantifiers offered by the epsilon terms has been utilised by proof assistants such as Isabelle and HOL \cite{AhrGie(1999.0)}. Epsilon terms also feature in philosophy and linguistics, and the reader interested in this is encouraged to consult the many references provided in \cite{AvZach(2002.0)}.

However, when it comes to concrete applications for the purpose of program extraction in mathematics or computer science, the epsilon calculus has been far superseded by the functional interpretations, which dominate this area for a number of reasons.

One property which the functional interpretations share is that the classical interpretations both factor through a natural intuitionistic counterpart. While the epsilon calculus applies directly to classical logic, the functional interpretations deal with this separately via the negative translation. For many applications of program extraction, particularly those which lean towards formal verification, the underlying proofs are indeed purely intuitionistic, in which case it makes far more sense to work with a direct implementation of the BHK interpretation such as modified realizability.

Another important feature of the functional interpretations is that the programs they extract are expressed as simple and clean lambda terms rather than an intricate transfinite recursion, as in the substitution method (though transfinite recursion is of course present in System T, it is conveniently encoded using the higher type recursors). In particular, lambda terms can be viewed in a very direct sense as functional programs, and are easily translated into real functional language like Haskell or ML. Therefore when it comes to the formal extraction of functional programs from intuitionistic proofs, variants of modified realizability are an obvious choice, and their success in this area is proven by the wide range of applications, ranging from real number computation \cite{BergMiySch(2016.0)} to the synthesis of SAT solvers \cite{BergLFS(2015.0)} (see the Minlog page at \cite{Minlog} for further examples).

For applications of proof interpretations in mathematics, in the sense of the \emph{proof mining} program \cite{Kohlenbach(2008.0)}, it is the Dialectica interpretation that has the starring role. Since applications of this kind typically deal with classical logic, they would in principle would also be suited to the epsilon calculus, and indeed early examples of proof mining made use of epsilon substitution \cite{Kreisel(1952.0)}. However, the fact that the Dialectica interpretation extracts functional programs is also crucial for modern proof mining, due to the direct use of \emph{mathematical properties} of those functionals. 

In proof mining, the Dialectica interpretation is typically used in its monotone variant, which technically speaking is a combination of the usual Dialectica interpretation with a bounding relation on functionals known as majorizability. It is precisely this combination of proof interpretation and majorizability which enables the extraction highly uniform, low complexity bounds from proofs which use what at first glance appear to be computationally intractable analytical principle such as Heine-Borel compactness. For much more detailed discussion on the role the Dialectica interpretation plays in proof mining, see \cite{Kohlenbach(2018.0)}. The main point to take away here is that the `functional' part of functional interpretations is a crucial part of their success!\bigskip

In the inevitably simplistic picture I have drawn above, I have deliberately contrasted the algorithmic elegance of the epsilon calculus with the applicability of the functional interpretations. A natural question is whether we can combine these two features in some way. This forms the main narrative of the remainder of the paper.

As we have seen, the functional interpretations deal with classical logic somewhat indirectly via negative translations. While in the case of realizability this can often be given a nice presentation in terms of games, in the case of the Dialectica interpretation in particular, the algorithm which is contained in the normalization of the term extracted from a negated proof is often very difficult to understand, and this is particularly the case once one moves away from simple examples like the drinkers paradox and analyzes `real' theorems from mathematics, where complex and abstract forms of recursion start to play a role.

Of the three interpretations, it is the Dialectica which has featured most prominently in my own research. I was drawn to it by its importance to the proof mining program, and the rich variety of mathematical theorems which have been studied using this interpretation, to which I also made a few small contributions.

In my own case studies \cite{OliPow(2015.1),OliPow(2015.0),OliPow(2017.0),Powell(2012.0),Powell(2013.0)} which focused on relatively strong theorems from mathematical analysis and well quasi-order theory, I was surprised by how the operational behaviour of the extracted terms were initially quite difficult to understand, but after carefully analysing them it became clear that they implemented rather clever backtracking algorithms. This led me to start thinking about the relationship between the Dialectica and the epsilon calculus, or more specifically, to ask whether terms extracted by the Dialectica in certain settings could be characterised as an epsilon-style procedures.

Although nowadays the epsilon calculus has become something of a specialised topic, in the past it was very much in the minds of those who pioneered applied proof theory. Early case studies by Kreisel were based on the substitution method, which in particular was used to prove soundness of the no-counterexample interpretation \cite{Kreisel(1951.0),Kreisel(1952.0)}, a close relative of the Dialectica interpretation (but see \cite{Kohlenbach(1999.0)}).

Particularly fascinating for me is the possibility that, while working on his groundbreaking extension of the Dialectica to full classical analysis, Spector was already thinking of a new way to deal with the axiom of choice using an epsilon-style algorithm. In his first footnote to \cite{Spector(1962.0)}, Kreisel writes of the draft of the paper he received from Spector before the latter's death:
\begin{quote}
\emph{This last half page states that the proof of the G\"{o}del translation of axiom F would use a generalization of Hilbert's substitution method as illustrated in the special case of \S12.1. However Spector's notes do not contain any details, so that it is not quite clear how to reconstruct the proof he had in mind.}
\end{quote}
Here axiom F is essentially the negative translation of the axiom of countable choice, which Spector had already interpreted using his schema of bar recursion in all finite types. While we do not know precisely what Spector intended, the idea of trying to replace the complicated forms of recursion which underlie the Dialectica with something more transparent is of great relevance now that proof interpretations are primarily used for practical program extraction as opposed to consistency proofs, and I talk about some more recent research in this direction in Section \ref{sec-dial-learning}.

The search for a clear algorithmic representation of terms extracted by the Dialectica led me to think, more generally, of how one could utilise concepts from the theory of programming languages to help capture what is going on `underneath the bonnet' of the Dialectica. To this end I worked on incorporating the notion of a global state into the interpretation via the state monad, and I discuss stateful programs in a much broader context in Section \ref{sec-dial-state}.

\section{Epsilon style algorithms and the Dialectica interpretation}
\label{sec-dial}

In Section \ref{sec-interpretations} I presented three old and well established computational interpretations of classical logic, by sketching how they act in the simple case of the drinkers paradox. My stress here was on the historical context in which they arose, and on highlighting key features of the interpretations which help explain the roles they play in modern proof theory. 

My goal in this section is to focus more closely on the connection between them, and more specifically, to demonstrate how the core idea underlying the epsilon calculus can help us better understand the Dialectica interpretation. As such, I consider two concepts in which epsilon substitution style algorithms can be elegantly phrased, but which are both much more general: Learning algorithms and stateful programs. 

I have used each of these to study the Dialectica interpretation, in \cite{Powell(2016.0)} and \cite{Powell(2018.1)} respectively, and these studies will form the main narrative of this section. Nevertheless, learning and programming with state feature much more broadly in modern approaches to program extraction, and I was certainly not the first to utilise them in this way. Therefore my priority here is to explain both in suitably general terms that their application in a broader context can be appreciated.

\subsection{Learning algorithms}
\label{sec-dial-learning}

As we have seen, the notion of learning as a computational semantics of classical logic in already explicitly present in the epsilon calculus, and can actually be seen in each of the interpretations of the drinkers paradox in Section \ref{sec-interpretations}. In fact, over the years this idea has continually resurfaced in a variety of different settings, some of which have already been mentioned (and a detailed survey of which would be an extensive work in its own right). 

In this section, I will introduce a specific form of learning phrased in the language of all finite types, which I found useful in clarifying certain aspects of the Dialectica interpretation in \cite{Powell(2016.0)}, characterising extracted programs as epsilon-style procedures. I will briefly sketch how this relates to the drinkers paradox, but here our running example is far less illuminating, and so I will go on to present a Skolem form $\DP^\omega$ of the drinkers paradox, which is solved using a simple kind of learning procedure that appears frequently in the literature.

Computational interpretations of classical logic via learning are particularly meaningful when one focuses on $\forall \exists \forall$-formulas. By a simple adaptation of our argument in Section \ref{sec-interpretations-dial} we see that the combination of the negative translation and dialectica interpretation carries out the following transformation on such formulas:
\begin{equation*}
\forall a\exists x\forall y Q(a,x,y)\mapsto \exists\Phi\forall a,g Q(a,\Phi ag,g(\Phi ag))
\end{equation*}
which happens to also coincide with Kreisel's no-counterexample interpretation for formulas of this type. Here, we can visualise $\Phi ag$ as constructing an approximation to the non-constructive object $x$ in the following precise sense: Rather than demanding some $x$ such that $Q(a,x,y)$ is valid for all $y$, we compute for any given $g$ an $x$ such that $Q(a,x,gx)$ holds. 

Such reformulations of $\forall\exists\forall$-theorems are widely studied in proof mining: In the case of convergence they go under the name of `metastability'. Here $x$ is a number, and one is typically concerned with producing a computable bound for the approximation of $x$ i.e. $\forall a,g\exists x\leq \Phi ag Q(a,x,gx)$ (see \cite{KohSaf(2014.0)} for an interesting discussion of this and several related concepts). However, $x$ could be a more complicated object, such as a maximal ideal or a choice sequence, and in this case approximations to $x$ are often built using a complex forms of higher-order recursion whose normalization is subtle and whose operational meaning can be difficult to intuit, but which can be cleanly and elegantly presented as learning procedures. 


So what is a learning procedure? In my own version of this well-known idea \cite{Powell(2016.0)}, I start with the idea of a learning \emph{algorithm}. A learning algorithm is assigned a type $\rho,\tau$, and, roughly speaking, seeks to build approximations to an object of type $\rho$ using building blocks of type $\tau$. Formally, it is given by a tuple $\mathcal{L}=(Q,\xi,\oplus)$ where the components are to be understood as follows:
\begin{itemize}

\item $Q$ is a predicate on $\rho$ which tells us if a suitably `good' approximation has been found;

\item $\xi:\rho\to \tau$ is a function which takes a current approximation $x$ and returns a new building block $\xi(x)$;

\item $\oplus:\rho\times \tau\to \rho$ combines $x$ with a building block $y$ to form a new approximation $x\oplus y$. 

\end{itemize}
Given any initial object $x_0$, a learning algorithm triggers a learning \emph{procedure} $x_0,x_1,x_2,\ldots$, which is defined recursively by
\begin{equation*}
x_{i+1}=\begin{cases}x_i & \mbox{if $Q(x_i)$}\\ x_i\oplus\xi(x_i) & \mbox{otherwise}\end{cases}
\end{equation*}
The basic idea is very simple. At each stage in the procedure we check whether or not $x_i$ is a good approximation. If it is, we leave it unchanged, if not, we assume that by its failure we have learned a new piece of information $\xi(x_i)$, which we can use to update the approximation $x_{i+1}=x_i\oplus\xi(x_i)$. We are typically interested in learning algorithms whose learning procedures terminate at some point, by which we mean there is some $k$ such that $Q(x_k)$ holds. In this case we say that $x_k$ is the \emph{limit} of the procedure and write $\lim\lp{x}:=x_k$.

The dialectica interpretation of the drinkers paradox can be solved by an extremely simple learning algorithm of length at most two: Given our counter function $g$ we define $\mathcal{L}$ by 
\begin{equation*}
Q(x)=P(x)\to P(gx) \ \ \ \ \ \ \xi=g \ \ \ \ \ \ x\oplus y=y
\end{equation*} 
and it is easy to see that $\Psi g:=\lim\lp{0}$ works. Of course this is essentially the same as the program $$\Phi g=\mbox{$0$ if $P(0)\to P(g0)$ else $g0$},$$ but here the learning that we see in the epsilon calculus is made explicit. For the sake of completeness, we give the summary of the algorithm in Figure 9.

\begin{figure}[h]
\begin{tcolorbox}


\textbf{Algorithm:} $\lim\lp{0}$
\[\xymatrix{& \fbox{$x_0:=0$}\ar[d] \\  & \fbox{Query: $P(x_0)\to P(gx_0)$}\ar[ld]_>>>>>>>>>{YES}\ar[d]^{NO} \\ \fbox{END} & \fbox{Update: $x_1:=x_0\oplus gx_0=gx_0$}\ar[l] }\]

\end{tcolorbox}
\caption{Interpretation of $\DP$ via learning procedure}
\end{figure}
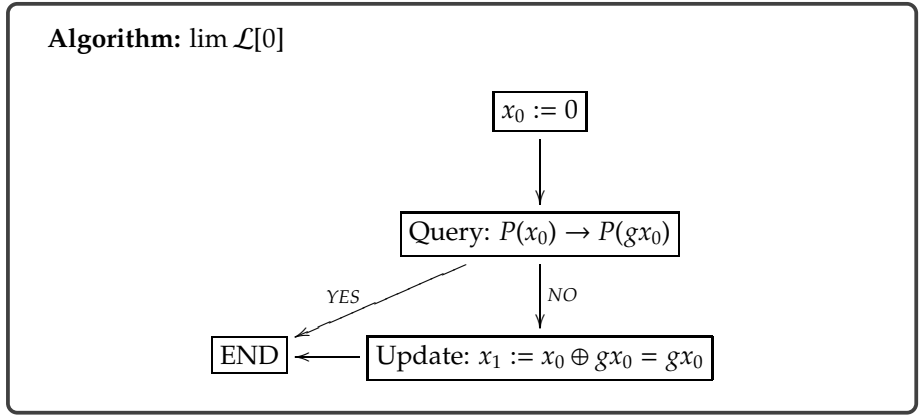

Unsurprisingly, learning algorithms in \cite{Powell(2016.0)} were not introduced to help better understand the drinkers paradox, but to give an algorithmic description of how complicated objects such as choice sequences are built, which is more or less the role they play elsewhere in the literature. The main part of my own paper is in fact the study of learning algorithms on infinite sequences, which are rather subtle and involve non-trivial termination arguments. 

To give a the reader a flavour of a meaningful learning procedure, let's take instead of a single predicate $P$ a sequence $(P_n)$ of decidable predicates, and consider the following sequential form of the drinkers paradox:
\begin{equation*}
\DP^\omega \ \ \ \colon \ \ \ \exists f^{\NN\to\NN} \forall n,m(P_n(fn)\to P_n(m))
\end{equation*}
which is provable using classical logic in conjunction with the axiom of countable choice. The Dialectica interpretation of the negative translation of this (before bringing the $\exists f$ out to the front) is given by
\begin{equation*}
\forall \omega,\phi\exists f(P_{\omega f}(f(\omega f))\to P_{\omega f}(\phi f)) 
\end{equation*}
where $\omega,\phi$ are functionals of type $(\NN\to\NN)\to\NN$. In other words, we need to build an approximation to the Skolem function $f$ which works, not for all $n$ and $y$, but just for $\omega f$ and $\phi f$. We can construct such an $f$ in $\omega,\phi$ by taking the limit of the learning algorithm $\mathcal{L}[\lambda i.0]$ for $\mathcal{L}$ of type $\NN\to\NN$, $\NN\times \NN$ given by

\begin{itemize}

\item $Q(f):=P_{\omega f}(f(\omega f))\to P_{\omega f}(\phi f)$

\item $\xi(f):=(\omega f,\phi f)$

\item $f\oplus (n,x):=f[n\mapsto x]$

\end{itemize}
where $f[n\mapsto x]$ is the function $f$ updated with the value $x$ at argument $n$. It's immediately clear that once the underlying learning procedure terminates with some $f$ satisfying $Q(f)$ then we're done, as this this is just the definition of a realizer for the functional interpretation of $\DP^\omega$. What is less clear is what the procedure does and why it terminates!

The learning procedure generates a sequence of functions
\begin{equation*}f_0,f_1,f_2,\ldots\end{equation*}
where $f_{i+1}=f_i[n_i\mapsto x_i]$ (we write $(n_i,x_i):=(\omega f_i,\phi f_i)$ for simplicity). Whenever our current attempt $f_i$ fails i.e. $\neg Q(f_i)$ holds, we know in particular that $\neg P_{n_i}(x_i)$, and so we have learned something about our predicates $P_i$. Updating $f_{i+1}=f[n_i\mapsto x_i]$ means that $f_{i+1}$ is now a valid choice sequence at point $n_i$, since $P_{n_i}(x_i)\to P_{n_i}(y)$ for all $y$. Looking at the procedure as a whole, it's clear that $f_{i+1}$ is in fact a valid choice sequence for all of $n_0,\ldots,n_i$, and so our learning procedure can be viewed as building an increasingly accurate approximation to the ideal choice function $f$. 

Now we need to impose a condition on $\omega$ and $\phi$, namely that they are \emph{continuous}, which means that they only look at a finite part of their input. This is not an unreasonable assumption, since all computable functionals in particular are continuous, but it nevertheless means that our learning procedure cannot be defined as a term of System T, which is a good indication of the fact that we are going beyond the usual soundness theorem for the Dialectica interpretation, since $\DP^\omega$ is not provable in Peano arithmetic. 

For each $n$, the sequence $f_0(n),f_1(n),f_2(n),\ldots$ changes at most once, and thus the functions $f_0,f_1,f_2,\ldots$ tend to some limit. By continuity of $\omega$, this means that $\omega f_0,\omega f_1,\omega f_2,\ldots$ eventually becomes constant, and so at some point $j$ we must have $n_{j+1}=\omega f_{j+1}\in\{n_0,\ldots,n_j\}$ and so in particular $\neg P_{n_{j+1}}(f_{j+1}(n_{j+1}))$, which implies $Q(f_{j+1})$, and hence termination of the procedure. A diagram of the learning procedure as a whole is given in Figure 10. 

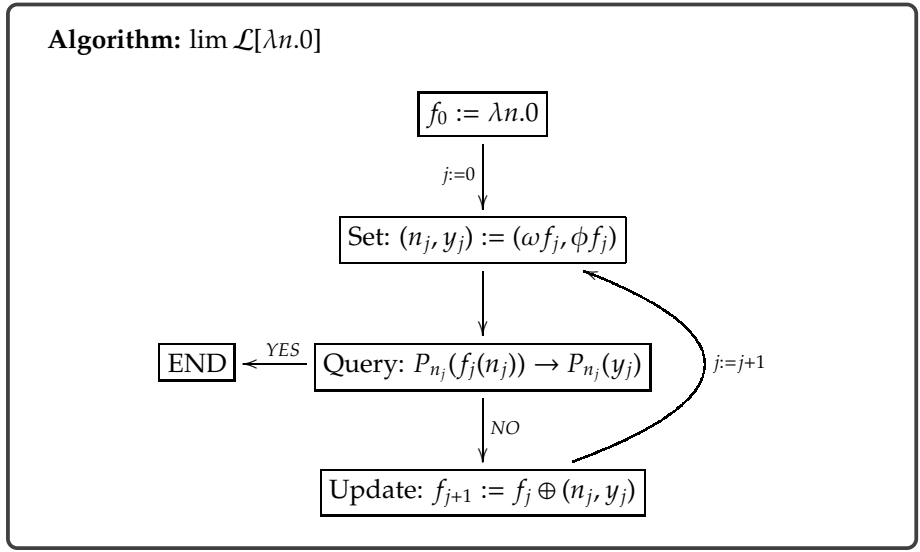
\begin{figure}[h]
\begin{tcolorbox}


\textbf{Algorithm:} $\lim\lp{\lambda n.0}$
\[\xymatrix{& \fbox{$f_0:=\lambda n.0$}\ar[d]_{j:=0} \\ & \fbox{Set: $(n_j,y_j):=(\omega f_j,\phi f_j)$}\ar[d] \\ \fbox{END} & \fbox{Query: $P_{n_j}(f_j(n_j))\to P_{n_j}(y_j)$}\ar[l]_>>>>>>{YES}\ar[d]^{NO} \\  & \fbox{Update: $f_{j+1}:=f_j\oplus (n_j,y_j)$}\ar@/_7pc/_{j:=j+1}[uu] }\]

\end{tcolorbox}
\caption{Interpretation of $\DP^\omega$ via learning procedure}
\end{figure}

The reader will hopefully have noticed that our learning procedure is nothing more than a simple version of epsilon substitution, formulated in the higher-order language of the Dialectica interpretation and used to produce a realizer for the Dialectica interpretation of a choice principle! We imagine the function $f$ in $\DP^\omega$ as finding some element such that $\neg P_i(m)$ holds, if it exists, and as such identify $fn$ with $\epsilon m\neg P_n(m)$. Our learning procedure starts by setting $fn=0$ for all $n$, and then, via a process of trial and error, repeatedly corrects the function until a sufficiently good approximation has been found. The current critical formula is given by $Q(f)$. 

The idea of solving $\DP^\omega$ in this way can already be found in \S12.1 of Spector \cite{Spector(1962.0)}, and it is precisely this that he claims to have generalised to solve the functional interpretation of full countable choice. While his proof was never recovered, learning procedures have since emerged in several places, and many works on this topic can be characterised as the study of more complicated versions of this idea.

My own work in this direction has been primarily concerned with the Dialectica interpretation itself and the development of Spector's idea. In \cite{Powell(2016.0)} I study a more general class of learning procedure, and show in particular that Spector's bar recursion can be seen as a kind of `forgetful' learning procedure. I extend this idea in \cite{Powell(2018.0)}, exploiting the connection established in \cite{Powell(2014.0)} between bar recursion and Berger's open recursion \cite{Berger(2004.0)}, and devise learning procedure which solves the Dialectica interpretation of a simple form of Zorn's lemma.

However, the relationship between classical logic and learning has also been developed by other authors in different contexts. 
To give a few examples: In \cite{BBC(1998.0)}, a variation of the algorithm above is captured in a realizability setting via a form of recursion in higher-types now known as the BBC functional, which has since been studied in greater depth in \cite{Berger(2004.0)}. In \cite{Avigad(2002.0)}, learning procedures are called \emph{update procedures}, and are used to prove the $1$-consistency of arithmetic. 

Finally, and perhaps most importantly, in \cite{AscBer(2010.0)} and later works by these authors (see in particular \cite{Aschieri(2011.1),Aschieri(2011.0)}), a special form of realizability is developed which is is based entirely on the notion of learning. Here, learning is captured through a `state of knowledge', which contains a finitary approximation to a Skolem function. Realizers for existential statements are evaluated relative to this state, and they either succeed, or discover an error in the current state, in which case the state is updated and the process repeats. In this, more than any of the other works mentioned here, learning is characterised as a stateful program, which brings me to the final part of the paper.
 

\subsection{Stateful programs}
\label{sec-dial-state}

Learning procedures are illuminating precisely because they capture the computational content of classical reasoning on an algorithmic level, through the \emph{evolution of a state}. The idea of a computation as an action on a state is the foundation of the imperative programming paradigm, on which many major programming languages - including C++, Java and Python - are partially based. This begs a much broader question: Can we utilise ideas from the theory of imperative languages to more clearly and elegantly express the computational meaning of classical reasoning?

This point of view has been explored primarily in the French style of program extraction via e.g. Krivine realizability, in which there is a greater tendency to capture the computational content of classical logic via control operators such as call/cc, rather than through logical techniques like negative translations (see \cite{Miquel(2011.0)} for an interesting discussion of this). 


However, my emphasis here is somewhat different: Namely on adapting \emph{traditional} proof interpretations like the Dialectica so that the operational behaviour of extracted programs can be better understood. The benefits for this are great: the Dialectica has undoubtedly the richest catalogue of applications of all proof interpretations, and so a characterisation of extracted terms as stateful programs could lead to fascinating connections between imperative languages and non-constructive principles from everyday mathematics.

This final section discusses ideas in this direction. My main goal will be to indicate how stateful programs can be simulated in a functional setting, and describe how they can be used to enrich the Dialectica interpretation and give it a more imperative flavour. Though this takes as its inspiration both the epsilon substitution and the learning procedures just presented, work of this kind is more general and focused on breaking down proof interpretations on a much lower level. It is also far less developed, with only a few recent works which view the traditional interpretations from this perspective, and as such this section can be seen as an extended conclusion, looking ahead to future research.\bigskip

I should start by clarifying the difference between imperative and functional programs. Simply put, imperative programs are built using commands which act on some underlying state. A while loop can be seen as a simple program written in an imperative style. Consider the following formulation of the factorial function:
\begin{equation*}
\begin{aligned}
&i:=1\\
&j:=1\\
&\text{\tt{while} $i< n$ do}\\
& \ \ \ \ \ \ i:=i+1\\
& \ \ \ \ \ \ j:=i\cdot j\\
&\text{\tt{print} $j$}
\end{aligned}
\end{equation*}
The loop is preceded by some variable assignments which determine an initial state, each iteration of the loop modifies the state until it terminates, and finally the output of the computation is read off from the state. 

Another program which fits the imperative paradigm is of course the epsilon substitution method, where one imagines epsilon terms as global variables: Our initial state assigns to each epsilon term the value $0$, and each stage of the method updates the state by correcting an epsilon term until a suitable approximation is found. The closely related learning procedures of the last section can be seen in a similar way.  

Programs extracted using the functional interpretations are, however, primarily expressed in some abstract functional language. In functional languages, the emphasis is on \emph{what} a program should do, rather than \emph{how} is does it. Programs are written as simple recursive equations and the concept of global state is not naturally present. In the functional style, the factorial function would be specified as just
\begin{equation*}
f(0)=1 \ \ \ \ \ \ \ \ f(n+1)=(n+1)\cdot f(n)
\end{equation*} 
The use of simple higher-order functional languages like System T is a central feature of the functional interpretations, and as highlighted in Section \ref{sec-interlude}, a key component of their success. Nevertheless, there \emph{is} a natural way of incorporating imperative features into functional calcui: The use a \emph{monad}, a structure familiar to any functional programmer. A full technical definition of a monad and it's role in programming is far beyond the scope of this paper (see e.g. \cite{Moggi(1991.0),Wadler(1995.0)}), but we will give quick sketch of the \emph{state monad}, which is of relevance in this section.



Suppose we are working in a simple functional calculus like System T and we want to capture some overriding global state which keeps track of certain aspects of the computations. To this end, we introduce to our calculus a new type $S$ of states, together with an mapping $T$ on types defined by
\begin{equation*}TX:=S\to X\times S.\end{equation*}
This is the state monad. Monads come equipped with two operators: A \emph{unit} of type $X\to TX$ and a \emph{bind} of type $TX\to (X\to TY)\to TY$. For the state monad these are given by the maps
\begin{equation*}
\begin{aligned}
\textrm{unit}(x)&:=\lambda s. (x,s)\\
\textrm{bind}(a)(b)&:=\lambda s. bxs'\mbox{ \ \ \ where $(x,s')=as$}
\end{aligned}
\end{equation*} 
In our enriched language, base types $X$ are interpreted as monadic types $TX=S\to X\times S$, while function types $X\to Y$ are interpreted as objects of type $X\to TY=X\to S\to Y\times S$. The unit map specifies a neutral translation from terms of plain type to the corresponding monadic type: For the state monad, neutrality means that the state remains unchanged. The bind map allows us to apply monadic functions to monadic arguments, as we will see below.

In the pure functional world, a term of ground type $X$ is just a program which evaluates to some value of that type. Under the state monad, it becomes a term which takes an initial state $s_1$ and returns a pair $(u,s_2)$ consisting of a value of type $X$ and a final state $s_2$. Similarly, in the pure word, a term of type $X\to Y$ is a function which takes an input $x$ and returns an output $y$. Under the state monad, it becomes a function which takes an input $x$ together with an initial state $s_1$ and returns an output-final state pair $(y,s_2)$. 

To emphasise this point, consider our two version of the factorial function. The purely functional one has type $\NN\to\NN$: it takes an input a number $n$ and evaluates to $n!$. The imperative version, however, acts on an underlying state: It takes as input the number $n$ together with the initial state $(i,j):=(1,1)$ and returns the output $n!$ together with the final state $(i,j):=(n,n!)$. In this sense it has type $\NN\to S\to \NN\times S$.

The bind operator is responsible for function application on the monadic level: It takes a monadic argument $a:TX$ together with a monadic function $b:X\to TY$ and returns a monadic output $\textrm{bind}(a)(b):TY$. This output mimics the following procedure: We take our initial state $s_1$ and plug it into our argument $a$ to obtain a pair $(u,s_2)$ of type $X\times S$. We treat $s_2$ as an intermediate state in the computation, and plug both it and the value $u$ into the monadic function $b$ to obtain a pair $(v,s_3)$ of type $Y\times S$. 

The purpose of all this is to try to convey to the reader how stateful functions like our while loop can be simulated in a functional environment via the state monad. We can, more generally, translate the finite types as a whole to a corresponding hierarchy of monadic types, and using the unit and bind operations define a translation on pure terms of System T to monadic terms in a variety of ways, depending on what kind of computation we are trying to simulate and what kind of information we aim to capture in our state. Again, this is presented in detail in  \cite{Moggi(1991.0),Wadler(1995.0)}.

So how can all this be applied to the Dialectica interpretation? Suppose we have a proof of a simple $\forall\exists $ statement $\forall x\exists y Q(x,y)$: The dialectica interpretation would normally extract a pure program
\begin{equation*}
f:X\to Y
\end{equation*}
which takes an input $x\in X$ and returns an output $y\in Y$ satisfying $Q(x,y)$. However, using the state monad together with a suitable translation on terms, we could instead set up the interpretation so that it extracts a state sensitive program 
\begin{equation*}
b:X\to S\to Y\times S
\end{equation*}
which takes an input $x\in X$ and initial state $s_1\in S$, and returns an output pair $bxs=(y,s_2)$, where $y$ satisfies $Q(x,y)$ and the final state $s_2$ tells us something about what our function \emph{did} during the computation, the idea being to make explicit in some sense the underlying substitution, or learning algorithm.

In a recent paper \cite{Powell(2018.1)}, I explore some of the things which a global state can do in this context, focusing in particular on the program's `interaction with the mathematical environment'. In other words, when performing a computation on some ambient mathematical structure, be it a predicate, a bounded monotone sequence, a colouring of a graph etc., the state monad is used to track the way in which the program queries this environment as it attempts to build an approximation to some object related to that structure e.g. an approximate drinker, an interval of metastability, an approximate monochromatic set etc. In this way, the state monad allows us to smoothly capture the epsilon style procedures which \emph{implicitly} underlie the normalization of extracted functional terms.

Let's try to illustrate this with a very simple example using, for the final time, the drinkers paradox. Note that the $\forall \exists$ formula we have in mind here is the (partial) Dialectica interpretation of $\DP^N$, namely $\forall g\exists x(P(x)\to P(gx))$. Suppose that our mathematical environment consists of a single predicate $P$, and that our states are defined to be finite piece of information about this predicate i.e. for $s\in S$ we have
\begin{equation*}
s=[Q_0(n_0),\ldots,Q_{k-1}(n_{k-1})]
\end{equation*}
for where $Q_i$ is either $P$ or $\neg P$. Our realizer could then take a state $s$ as an input, and whenever it tests $P$ on a given value it will add this information to the state. For the usual drinkers paradox, we would get the following variation of the program given in Section \ref{sec-dial}, which now has the monadic type $(\NN\to\NN)\to S\to \NN\times S$:
\begin{equation*}
\Phi gs:=\begin{cases}\pair{0,s::\neg P(0)} & \mbox{if $\neg P(0)$}\\
\pair{0,s:: P(0):: P(g0)} & \mbox{if $P(g0)$}\\
\pair{g0,s::P(0):: \neg P(g0)} & \mbox{otherwise}\end{cases}
\end{equation*}
This breaks our usual case distinction into parts, starting by testing the truth value of $P(0)$. If this is false, we know that $P(0)\to P(g0)$ is true and thus $0$ is realizer for $\DP$, and we add what we have learned about $P$ to the state. On the other hand, if $P(0)$ is true, the truth value of $P(0)\to P(g0)$ is still undetermined, so we must now test $P(g0)$, leading to two possible outcomes. Unlike the pure program, our stateful program returns a \emph{reason} as to why the witness it has returned works, giving us the information about the mathematical environment which has determined its choice. Alternatively put, our program returns a record of the underlying substitution procedure carried out by the term. As usual we present this as an algorithm in Figure 11. 

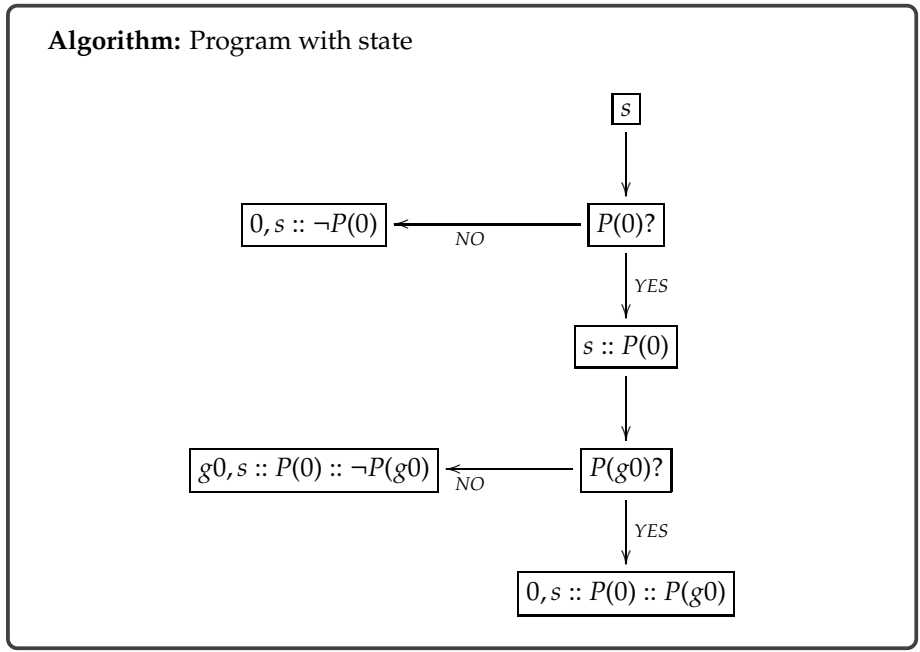
\begin{figure}[h]
\begin{tcolorbox}


\textbf{Algorithm:} Program with state
\[\xymatrix{& \fbox{$s$}\ar[d] \\ \fbox{$0,s::\neg P(0)$} & \fbox{$P(0)?$}\ar[l]^{NO}\ar[d]^{YES} \\ & \fbox{$s::P(0)$}\ar[d] \\ \fbox{$g0,s::P(0)::\neg P(g0)$} & \fbox{$P(g0)?$}\ar[l]^{NO}\ar[d]^{YES} \\ & \fbox{$0,s::P(0)::P(g0)$} }\]

%

\end{tcolorbox}
\caption{Interpretation of $\DP$ with state}
\end{figure}

We emphasise that this is just a simple illustration of how the state could be used to enhance the programs we extract - to just record the interactions which have taken place with the mathematical environment. In particular, the realizer makes no use of what is currently written in the state. One could change this so that it first looks to see whether a truth value of e.g. $P(0)$ is already present in $s$, and only then proceed to test $P(0)$. In cases where the cost of evaluating $P$ is high, this would improve the efficiency of the extracted program, allowing us to store previously computed values so that they can be accessed at a later stage in the computation. More sophisticated variants are possible: Where $P$ is not decidable, we can simply make an arbitrary choice for its truth value and make the validity of our output witness dependent on the truth of the state. In \cite{Powell(2018.1)} a very general framework for working with state is presented via an abstract soundness theorem, and a range of applications are discussed, ranging from Herbrand's theorem to program synthesis. 

I am certainly not alone in turning towards concepts such as monads to try to better explain what is going on under the syntax of traditional interpretations, although to date there is comparatively little research in this direction. Similar work, also pertaining to the Dialectica interpretation, has been carried out by Pedrot in \cite{Pedrot(2015.0)}. Berger et al. have used the state monad to automatically extract list sorting programs using modified realizability \cite{BergSeiWoo(2014.0)}, while the state monad is utilised by Birolo in \cite{Birolo(2012.0)} to give a more general version of Aschieri-Berardi learning realizability. Interestingly, monads also feature prominently in the construction of the Dialectica categories \cite{dePaiva(1991.0)} - abstract presentations of the interpretation which formed early models of linear logic - though there they play a somewhat different role.

\section{Conclusion}
\label{sec-conc}

I began this paper with an introduction to the epsilon calculus, a proof theoretic technique almost a century old and originating in the foundational crisis of mathematics. I concluded in the present day by sketching how stateful programs are being used to capture algorithmic aspects of proof interpretations and characterise the operational behaviour of extracted programs. 

Yet the ideas which underpin the latter are in some way already present in the substitution method, illustrating how certain universal characteristics seem to underlie computational interpretations of classical logic, manifesting themselves in different ways in different settings.


In my short presentations of the three traditional proof interpretations, I hope to have hinted at the deep connections that they share, which are not necessarily apparent in their formal definitions. Some of these connections have been made more explicit in modern research, in particular the utilisation of epsilon substitution-like procedures in the setting of functional interpretations, which formed the main topic of the second part of this paper.

As proof theory moves away from foundational concerns and towards applications in mathematics and computer science, is has become increasingly relevant to utilise new languages and techniques to modernise traditional proof theoretic methods, and it will be fascinating to see how proof interpretations continue to evolve.




\bibliographystyle{plain}
\bibliography{/home/thomas/Dropbox/tp}
\end{document}